\documentclass[12pt]{article}

\usepackage[dvips]{graphicx}

\makeatletter
\def\@cite#1#2{{[{#1}]\if@tempswa\typeout
{IJCGA warning: optional citation argument
ignored: `#2'} \fi}}
\newcount\@tempcntc
\def\@citex[#1]#2{\if@filesw\immediate\write\@auxout{\string\citation{#2}}\fi
  \@tempcnta\z@\@tempcntb\m@ne\def\@citea{}\@cite{\@for\@citeb:=#2\do
    {\@ifundefined
       {b@\@citeb}{\@citeo\@tempcntb\m@ne\@citea\def\@citea{,}{\bf ?}\@warning
       {Citation `\@citeb' on page \thepage \space undefined}}%
    {\setbox\z@\hbox{\global\@tempcntc0\csname b@\@citeb\endcsname\relax}%
     \ifnum\@tempcntc=\z@ \@citeo\@tempcntb\m@ne
       \@citea\def\@citea{,}\hbox{\csname b@\@citeb\endcsname}%
     \else
      \advance\@tempcntb\@ne
      \ifnum\@tempcntb=\@tempcntc
      \else\advance\@tempcntb\m@ne\@citeo
      \@tempcnta\@tempcntc\@tempcntb\@tempcntc\fi\fi}}\@citeo}{#1}}
\def\@citeo{\ifnum\@tempcnta>\@tempcntb\else\@citea\def\@citea{,}%
  \ifnum\@tempcnta=\@tempcntb\the\@tempcnta\else
   {\advance\@tempcnta\@ne\ifnum\@tempcnta=\@tempcntb \else \def\@citea{--}\fi
    \advance\@tempcnta\m@ne\the\@tempcnta\@citea\the\@tempcntb}\fi\fi}
\makeatother
\newenvironment{Eqnarray}%
     {\arraycolsep 0.14em\begin{eqnarray}}{\end{eqnarray}}

\def\be{\begin{equation}}
\def\ee{\end{equation}}
\def\bear{\be\begin{array}}
\def\eear{\end{array}\ee}
\def\bea{\begin{Eqnarray}}
\def\eea{\end{Eqnarray}}

\def\lsim{\mathrel{\raise.3ex\hbox{$<$\kern-.75em\lower1ex\hbox{$\sim$}}}}
\def\gsim{\mathrel{\raise.3ex\hbox{$>$\kern-.75em\lower1ex\hbox{$\sim$}}}}
\def\ifmath#1{\relax\ifmmode #1\else $#1$\fi}
\def\ls#1{\ifmath{_{\lower1.5pt\hbox{$\scriptstyle #1$}}}}

\def\beq{\begin{equation}}
\def\eeq{\end{equation}}
\def\beqa{\begin{Eqnarray}}
\def\eeqa{\end{Eqnarray}}

\def\boxit#1{\leavevmode\thinspace\hbox{\vrule\vtop{\vbox{\hrule%
        \vskip3pt\kern1pt\hbox{\vphantom{\bf/}\thinspace\thinspace%
        {\bf#1}\thinspace\thinspace}}\kern1pt\vskip3pt\hrule}\vrule}%
        \thinspace}
\def\Boxit#1{\noindent\vbox{\hrule\hbox{\vrule\kern3pt\vbox{
        \advance\hsize-7pt\vskip-\parskip\kern3pt\bf#1
        \hbox{\vrule height0pt depth\dp\strutbox width0pt}
        \kern3pt}\kern3pt\vrule}\hrule}}

\def\baselinestretch{1}
\begin{document}

%%%%%%%%%%%%%%%%%%%%%%%%%%% subequations.sty %%%%%%%%%%%%%%%%%%%%%%%%
\catcode`@=11
\newtoks\@stequation
\def\subequations{\refstepcounter{equation}%
\edef\@savedequation{\the\c@equation}%
  \@stequation=\expandafter{\theequation}%   %only want \theequation
  \edef\@savedtheequation{\the\@stequation}% % expanded once
  \edef\oldtheequation{\theequation}%
  \setcounter{equation}{0}%
  \def\theequation{\oldtheequation\alph{equation}}}
\def\endsubequations{\setcounter{equation}{\@savedequation}%
  \@stequation=\expandafter{\@savedtheequation}%
  \edef\theequation{\the\@stequation}\global\@ignoretrue
\noindent}
\catcode`@=12
%%%%%%%%%%%%%%%%%%%%%%%%%%%%%%%%%%%%%%%%%%%%%%%%%%%%%%%%%%%%%%%%%%%%%
\setcounter{footnote}{1} \setcounter{page}{1}

\noindent

\title{{\bf Tachyonic Squarks in Split Supersymmetry}}
\vskip2in
\author{    
{\bf Alejandro Ibarra\footnote{\baselineskip=16pt  E-mail: {\tt
alejandro.ibarra@cern.ch}}} \\ 
\hspace{3cm}\\
%\vskip.35in   
{\small Instituto de F\'{\i}sica Te\'orica, CSIC/UAM, C-XVI} \\
{\small Universidad Aut\'onoma de Madrid,} \\
{\small Cantoblanco, 28049 Madrid, Spain.}
}
\maketitle

\def\baselinestretch{1.15}
\begin{abstract}
\noindent
The decoupling of scalar particles in split supersymmetry
makes the spectrum of squarks irrelevant for low
energy processes. Nevertheless, the structure of the vacuum
is sensitive to the spectrum of squarks, even when
the supersymmetry breaking scale is large. In this note,
we show that in certain regions of the parameter space, 
squarks could develop radiatively tachyonic masses,
thus breaking electric charge and color. We discuss the constraints
that follow from the requirement of charge and color
conservation, and we comment on the implications for
model building.
\end{abstract}

\thispagestyle{empty}

\vskip-14.5cm
\rightline{IFT-UAM/CSIC-05-18}

%%%%%%%%%%%%%%%%%%%%%%%%%%%%%%%%%%%%%%%%%%%%%%%%%%%%%%%%%%%%%%%%%%%
\newpage
\baselineskip=20pt

\section{Introduction}

Low energy supersymmetry stands since many years as the most
attractive extension of the Standard Model. It provides not only
a theoretically well motivated solution to the hierarchy problem,
but also predicts the unification of the gauge couplings at a high
energy scale \cite{Dimopoulos:1981yj} and provides 
a promising dark matter candidate, the neutralino \cite{DM}. 
Despite the great interest of the supersymmetric extension of the
Standard Model, it is not free of problems. It suffers from
too large contributions to flavour changing 
neutral currents, CP violation and proton decay. 
Nevertheless, it is possible to circumvent all these problems
by adjusting the parameters of the model, being the most
simple solution to assume that squark and slepton masses are
sufficiently large, at least for the first two generations.
Above all, the most important drawback for the Supersymmetric Standard
Model is the failure in the quest of the Higgs boson, 
predicted to be fairly light in its minimal version, the MSSM.
To satisfy the experimental constraints, soft masses in the Higgs
sector have to be somewhat larger than the electroweak scale,
introducing a milder hierarchy problem.

This milder version of the naturalness problem could be interpreted as
an indication for physics beyond the Minimal Supersymmetric 
Standard Model. It could be alleviated, for instance,
by extending the model with one extra singlet, 
solving at the same time the $\mu$-problem \cite{Bastero-Gil:2000bw}. 
A more radical attitude to the naturalness 
problem was recently advocated by Arkani-Hamed
and Dimopoulos \cite{Arkani-Hamed:2004fb}, and consists 
in just accepting a fine-tuning in
the breaking of the electroweak symmetry, arguing
that there already exists a second (and more severe) hierarchy problem
in the Supersymmetric Standard Model, namely the cosmological
constant problem. With this guiding principle, there is no
reason to keep the scalar particles light, as long as
an (unspecified) mechanism can fine tune the Higgs 
vacuum expectation value to 246 GeV and the cosmological constant
to $\sim (10^{-3} ~{\rm eV})^4$.

The authors in 
\cite{Arkani-Hamed:2004fb,Giudice:2004tc,Arkani-Hamed:2004yi}
 also noted that making squarks and sleptons
heavy provides a solution to the problems of
Supersymmetric Standard Model, but does not necessarily
destroy the successes. Keeping the gauginos and higgsinos
at the electroweak scale, gauge unification is preserved
and the neutralino is still a viable candidate for the dark
matter of the Universe. Following Giudice and Romanino,
we will call this scenario split supersymmetry: a
scenario with light fermion masses and heavy scalar masses, 
except for the Standard Model Higgs,
fine tuned to yield a correct $Z$ boson mass and a small
cosmological constant. Recently, some of the low energy
implications of this scenario have been discussed,
such as electric dipole moments \cite{Arkani-Hamed:2004yi,EDM}, 
collider signatures \cite{Giudice:2004tc,collider},
Higgs physics and the electroweak symmetry breaking 
\cite{Giudice:2004tc,higgs,EW}, dark matter 
\cite{Giudice:2004tc,Arkani-Hamed:2004yi,DM-split}
or cosmic ray showers \cite{Anchordoqui:2004bd}.

The decoupling of the scalar particles might suggest that
the spectrum of squarks and sleptons is completely irrelevant
for the low-energy phenomenology. In this note, we would like
to point out that the structure of the supersymmetric vacuum
is indeed sensitive to the spectrum of squarks and sleptons,
even when the scale of supersymmetry breaking is large.
We will show that under certain conditions,
radiative corrections could induce tachyonic stop masses, 
thus leading to charge and colour breaking.
The reason for this can be easily understood from
the well known mechanism of radiative electroweak 
symmetry breaking in the MSSM \cite{Ibanez:1982fr}. 
Radiative corrections from the top Yukawa coupling
can drive the up-type Higgs mass squared to negative
values, thus breaking $SU(2)_L\times U(1)_Y$. This is not
normally the case for the stop mass squared, since
the gluino radiative corrections induce a 
positive contribution to the mass squared 
that is usually large enough to 
keep the stop mass squared positive. 
In split supersymmetry the gluino mass is much smaller
than the scalar masses, so this positive contribution is no
longer important, and in consequence there exists the possibility
of generating radiatively tachyonic stop masses. In this note
we will discuss the constraints that this imposes on the
scenario of split supersymmetry.

\section{Running of the squark masses}

Let us consider first scenarios with low $\tan\beta$, so 
that only the top Yukawa coupling is relevant. Later on, we will 
discuss scenarios with large $\tan\beta$ for which the effects 
from the bottom and tau Yukawa couplings also have to be taken into
account.
The one loop renormalization group equations for the
left and right handed stops and the up-type Higgs doublet
read:
\bea
&&\frac{dm_{\widetilde t_L}^2}{dt}=\frac{1}{16\pi^2} 
\left[\frac{16}{3} g^2_3 M^2_3
+3 g^2_2 M^2_2+\frac{1}{1 5} g^2_1 M^2_1
- h_t^2 (m_{\widetilde t_L}^2+m_{\widetilde t_R}^2+m_{H_u}^2+A_t^2) \right],
\nonumber\\
&&\frac{dm_{\widetilde t_R}^2}{dt}=\frac{1}{16\pi^2} 
\left[\frac{16}{3} g^2_3 M^2_3+\frac{16}{15} g^2_1 M^2_1-
2 h_t^2 (m_{\widetilde t_L}^2+m_{\widetilde t_R}^2+m_{H_u}^2+A_t^2)  \right],
\nonumber\\
&&\frac{dm_{H_u}^2}{dt}=\frac{1}{16\pi^2} 
\left[3 g^2_2 M^2_2+\frac{3}{5} g^2_1 M^2_1
-3 h_t^2 (m_{\widetilde t_L}^2+m_{\widetilde t_R}^2+m_{H_u}^2+A_t^2)  \right],
\label{RGEs}
\eea
where $t=\log(M_{X}^2/Q^2)$, $Q$ is the renormalization
scale and $M_{X}$ is the scale at which the soft terms are generated,
that we use as boundary condition to run the renormalization group equations. 
Inspired by the gravity mediated supersymmetry breaking 
framework, we will assume that
the soft breaking terms are generated at the reduced Planck scale,
$M_P=M_{Planck}/\sqrt{8 \pi}=2.4 \times 10^{18}$ GeV.

The set of differential equations (\ref{RGEs}) can be solved
analytically \cite{Ibanez:1984vq,Lleyda:1993xf}. 
In the limit of split supersymmetry,
gaugino masses are much smaller than the scalar masses,
and the trilinear soft terms, being protected by the
same R-symmetry that protects gaugino masses, are also expected
to be much smaller than the scalar masses. Therefore, the
analytical expression for the solution greatly simplifies,
and reads:
\bea
&&m_{\widetilde t_L}^2(t) =  m_{\widetilde t_L}^2(0) 
-\frac{h_t^2(0)}{(4 \pi)^2} \frac{F(t)}{D(t)}
[ m_{H_u}^2(0)+m_{\widetilde t_R}^2(0)+m_{\widetilde t_L}^2(0)], \nonumber \\
&&m_{\widetilde t_R}^2(t) =  m_{\widetilde t_R}^2(0) 
-2\frac{h_t^2(0)}{(4 \pi)^2} \frac{F(t)}{D(t)}
[m_{H_u}^2(0)+m_{\widetilde t_R}^2(0)+m_{\widetilde t_L}^2(0)], \nonumber \\
&&m_{H_u}^2(t) =  m_{H_u}^2(0) -3\frac{h_t^2(0)}{(4 \pi)^2} \frac{F(t)}{D(t)}
[m_{H_u}^2(0)+m_{\widetilde t_R}^2(0)+m_{\widetilde t_L}^2(0)],
\label{solution}
\eea
where
\bea
F(t)&\equiv& \int_0^t E(t') d t', \\
D(t)&\equiv& 1+ 6\frac{h_t^2(0)}{16\pi^2} F(t)
\label{defs-FD}
\eea
and
\bea
E(t)&\equiv& (1+ \frac{g^2_3(0)}{16\pi^2} b_3 t)^{(16/{3 b_3})}~
(1+ \frac{g^2_2(0)}{16\pi^2} b_2 t)^{(3/{ b_2})}
~(1+\frac{g^2_1(0)}{16\pi^2} b_1  t)^{(13/{15 b_1})},  
\label{def-E}
\eea
being $(b_1,b_2,b_3) \equiv (33/5,1,-3)$ the coefficients of the
beta functions for the gauge couplings. 

If the left-handed or the right-handed stop mass squared are driven
to negative values before they decouple, the corresponding field
will acquire a vacuum expectation value, thus yielding
a vacuum where charge and colour are not conserved. 
This situation is clearly undesirable and to prevent it
one has to require  $m^2_{\widetilde t_L}(t_{\widetilde m})> 0$
and $m^2_{\widetilde t_R}(t_{\widetilde m})> 0$, 
where $t_{\widetilde m}= \log(M^2_{X}/{\widetilde m}^2)$
and $\widetilde m$ is the typical size of the soft terms.
These conditions translate into constraints on the scalar
mass spectrum, that otherwise is completely unconstrained
by low energy experiments.

For definiteness, let us discuss first the limit 
in which the top Yukawa coupling is barely perturbative 
at the cut-off scale, {\it i.e.} the
infrared fixed point scenario, recently revisited
for the framework of split supersymmetry in \cite{Huitu:2005ef}.
In this limit, squark masses quickly reach the
fixed point (as long as $\widetilde m\lsim 10^{16}$ GeV),
usually before the squarks decouple.
The corresponding masses can be read from eqs.(\ref{solution})
by taking the limit $h_t(0) \rightarrow \infty$, or by
substituting
\bea
\frac{h_t^2(0)}{(4 \pi)^2} \frac{F(t)}{D(t)} \simeq \frac{1}{6}.
\eea
The result is
\bea
&&m^2_{\widetilde t_L}(t_{\widetilde m}) \simeq \frac{1}{6}[
5\,m_{\widetilde t_L}^2(0)-m_{H_u}^2(0)-m_{\widetilde t_R}^2(0)], \nonumber \\
&&m^2_{\widetilde t_R}(t_{\widetilde m}) \simeq \frac{1}{3}
[2\,m_{\widetilde t_R}^2(0)-m_{H_u}^2(0)-m_{\widetilde t_L}^2(0)],\nonumber \\
&&m^2_{H_u}(t_{\widetilde m}) \simeq \frac{1}{2}[
m_{H_u}^2(0)-m_{\widetilde t_L}^2(0)-m_{\widetilde t_R}^2(0)].
\label{conditions-IFP}
\eea

\begin{figure}[t!]%%%%%%%%%%%%%%%%%%%%%%%%%%%%%%%%%%%%%%%%%%%
  %%%
  \centerline{
    \scalebox{0.45}{\includegraphics{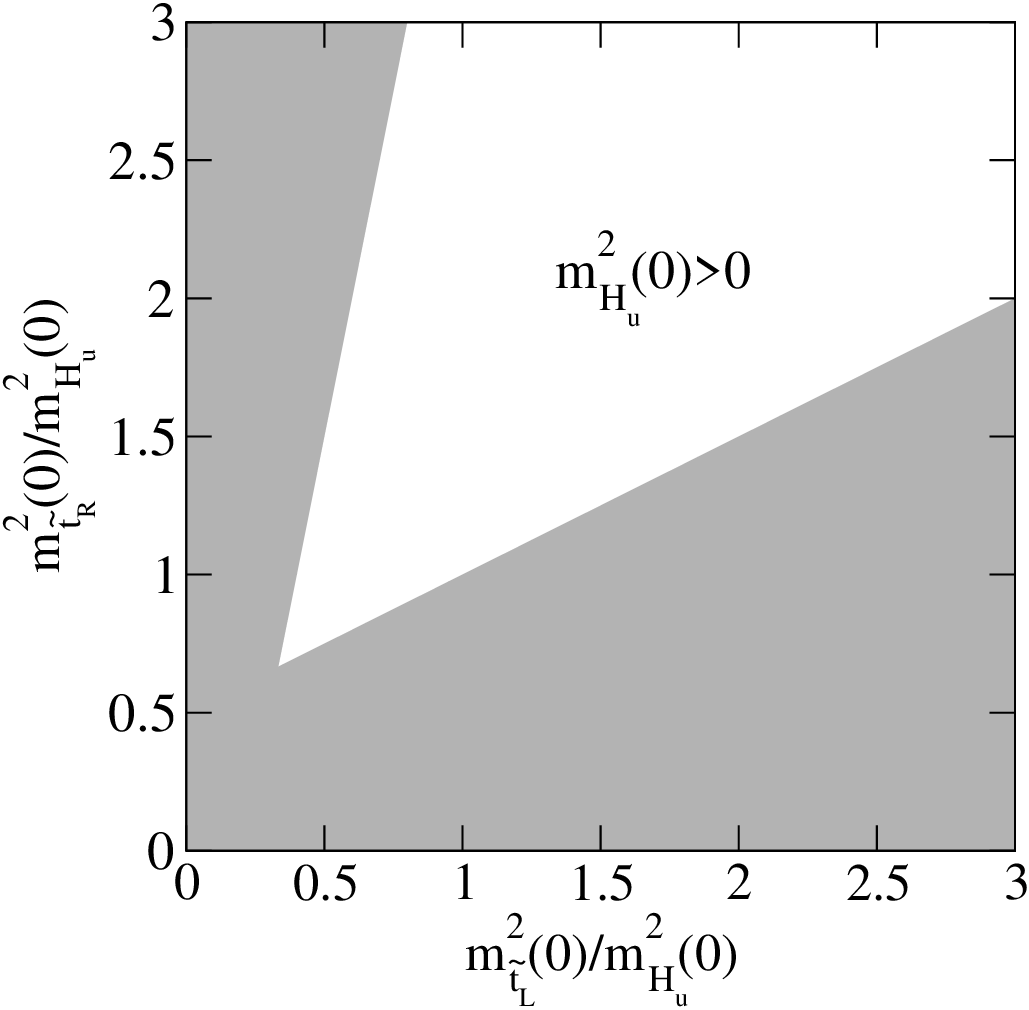}}~~~
    \scalebox{0.45}{\includegraphics{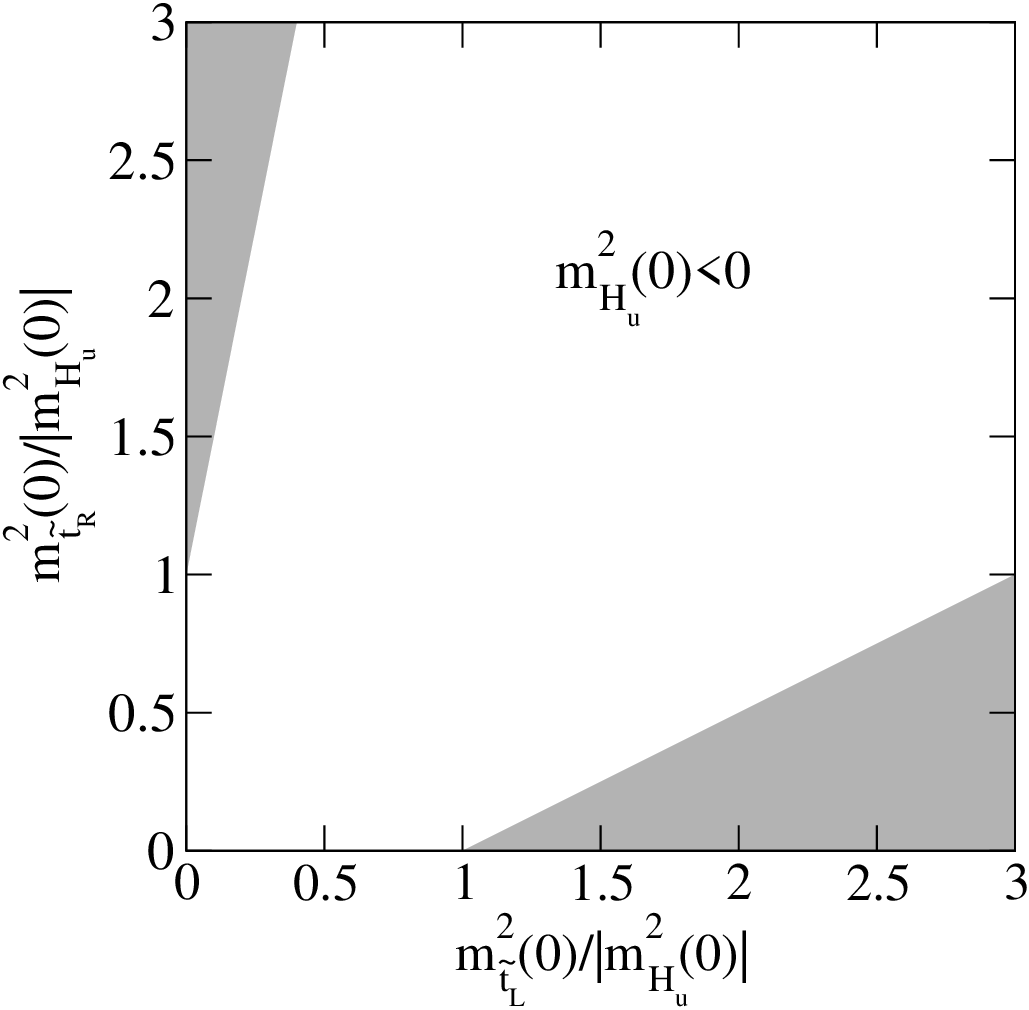}}
  }
  \caption{Allowed region in the stop parameter space
from imposing the requirement of charge and colour conservation,
in the infrared fixed point limit for the top Yukawa coupling.
In the left plot, it is shown the case in which the up-type
Higgs mass squared is positive at the cut-off scale, whereas in
the right plot we assume that it is negative.}
  \label{fig-IFP}
\end{figure}

The region of the parameter space allowed from the requirement
of charge and colour conservation is shown in Fig \ref{fig-IFP}.
The parameter space is span by the left and right handed 
stop masses squared relative to 
the up-type Higgs mass squared, whose absolute value
we assume equal to $\widetilde m^2$.
In the left plot we have assumed that the up-type Higgs mass
squared is positive at the reduced Planck scale, so that the electroweak
symmetry breaking is triggered by the familiar
mechanism of dimensional transmutation. For this 
to happen, the soft masses
at the high energy scale should satisfy the constraint
$m_{H_u}^2(0)-m_{\widetilde t_L}^2(0)-m_{\widetilde t_R}^2(0) < 0$.
Notice that the requirement of positive stop masses
squared is sufficient to guarantee the radiative 
breaking of the electroweak symmetry. In this case,
we find that large areas in the region with $m^2_{H_u}\gsim 
m^2_{\widetilde t_L, \widetilde t_R}$ are in conflict
with the requirement of charge and colour conservation.
On the other hand,
if there is a mechanism that will eventually fine-tune
the $Z$ mass to a small value, one cannot exclude
the possibility that the electroweak symmetry breaking
is taking place already at $M_P$. It could happen that 
the breaking of supersymmetry gives rise
to tachyonic up-type Higgs masses, thus leading
to the breaking of the electroweak symmetry 
at tree level. In the case that the Higgs mass is 
already tachyonic to start with, 
the  conditions $m^2_{\widetilde t_L}(t_{\widetilde m})> 0$
and $m^2_{\widetilde t_R}(t_{\widetilde m})> 0$ 
would be easier to fulfill, as can be realized 
from Fig.\ref{fig-IFP}, right plot.

%%%%%%%%%%%%%%%%%%%%%%%%%%%%%%%%%%%%%%%%%%%%%%%%%%%%%%%%%%%%
%%%%%%%%%%%%%%%%%%%%%%%%%%%%%%%%%%%%%%%%%%%%%%%%%%%%%%%%%%%

In a scenario with strict universality at the high energy scale,
stops will not develop tachyonic masses. 
Nevertheless, the large mass splitting between the
left-handed and the right-handed stops, that
belong to the same 10-plet of $SU(5)$, would spoil
the successful prediction for gauge unification.

\begin{figure}[t!]%%%%%%%%%%%%%%%%%%%%%%%%%%%%%%%%%%%%%%%%%%%
  %%%
  \centerline{
    \scalebox{0.45}{\includegraphics{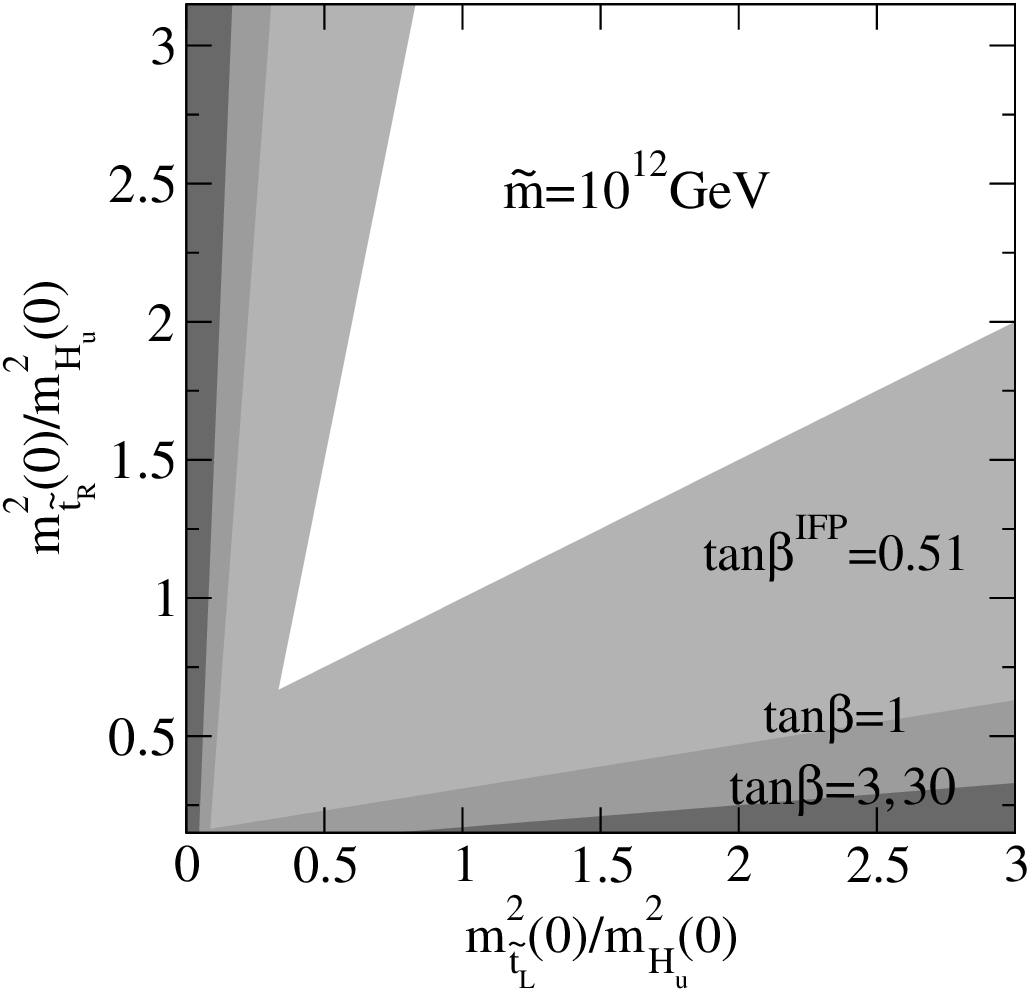}}~~~
    \scalebox{0.45}{\includegraphics{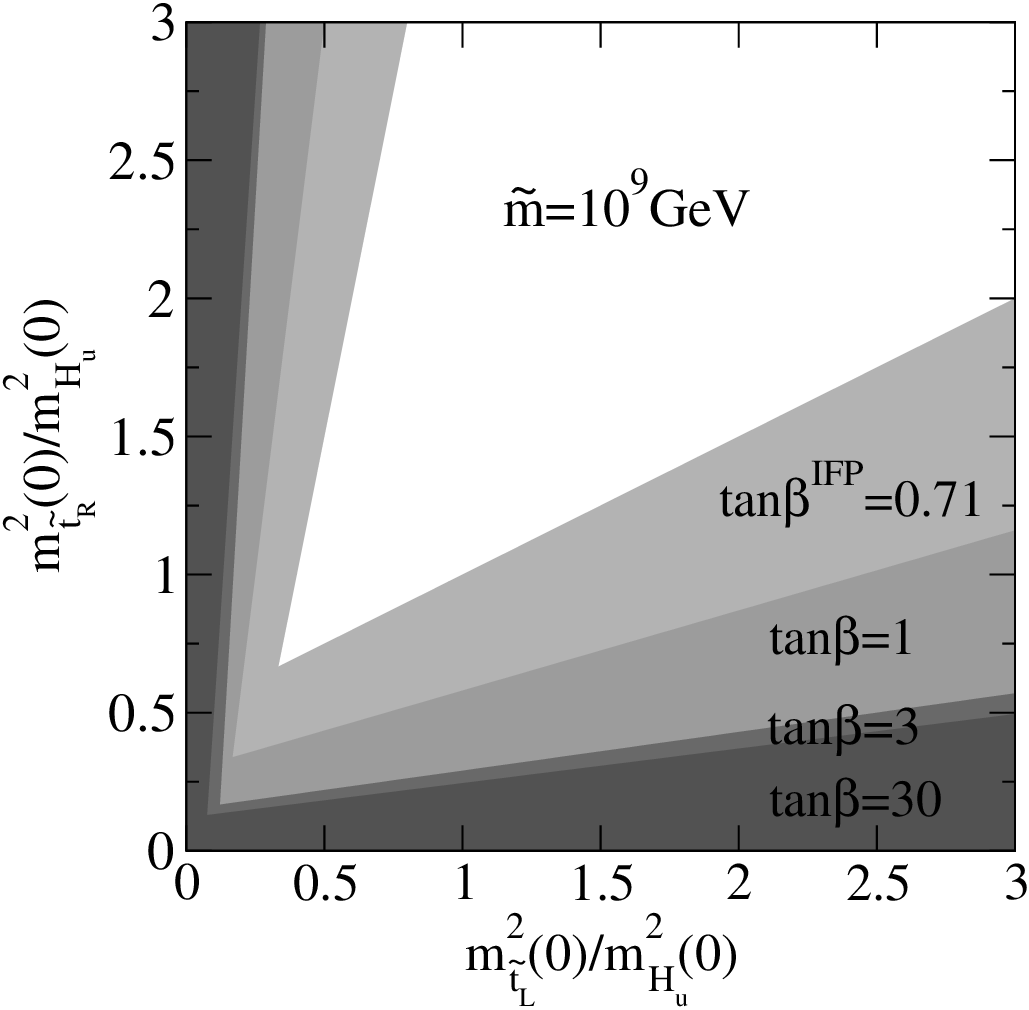}}} 
  \vspace{1.3cm}
  \centerline{
    \scalebox{0.45}{\includegraphics{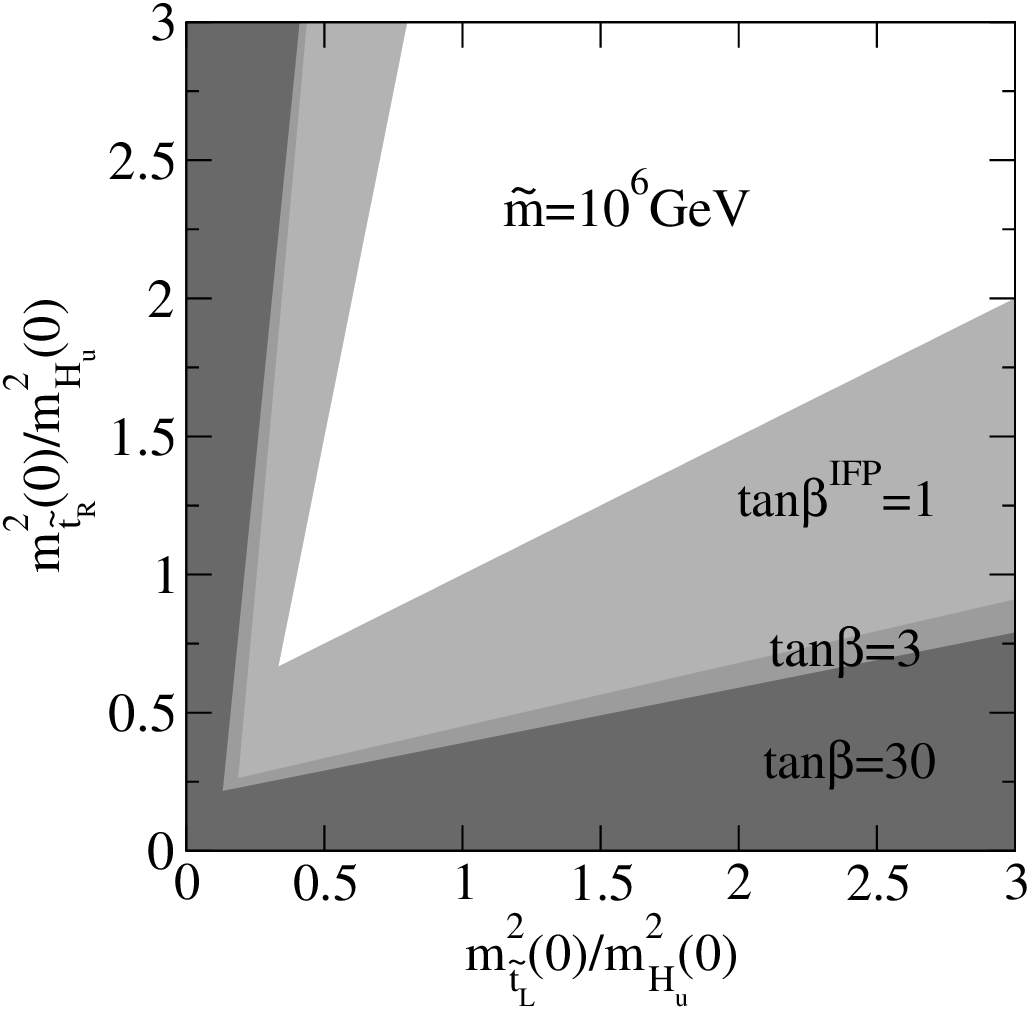}}~~~
    \scalebox{0.45}{\includegraphics{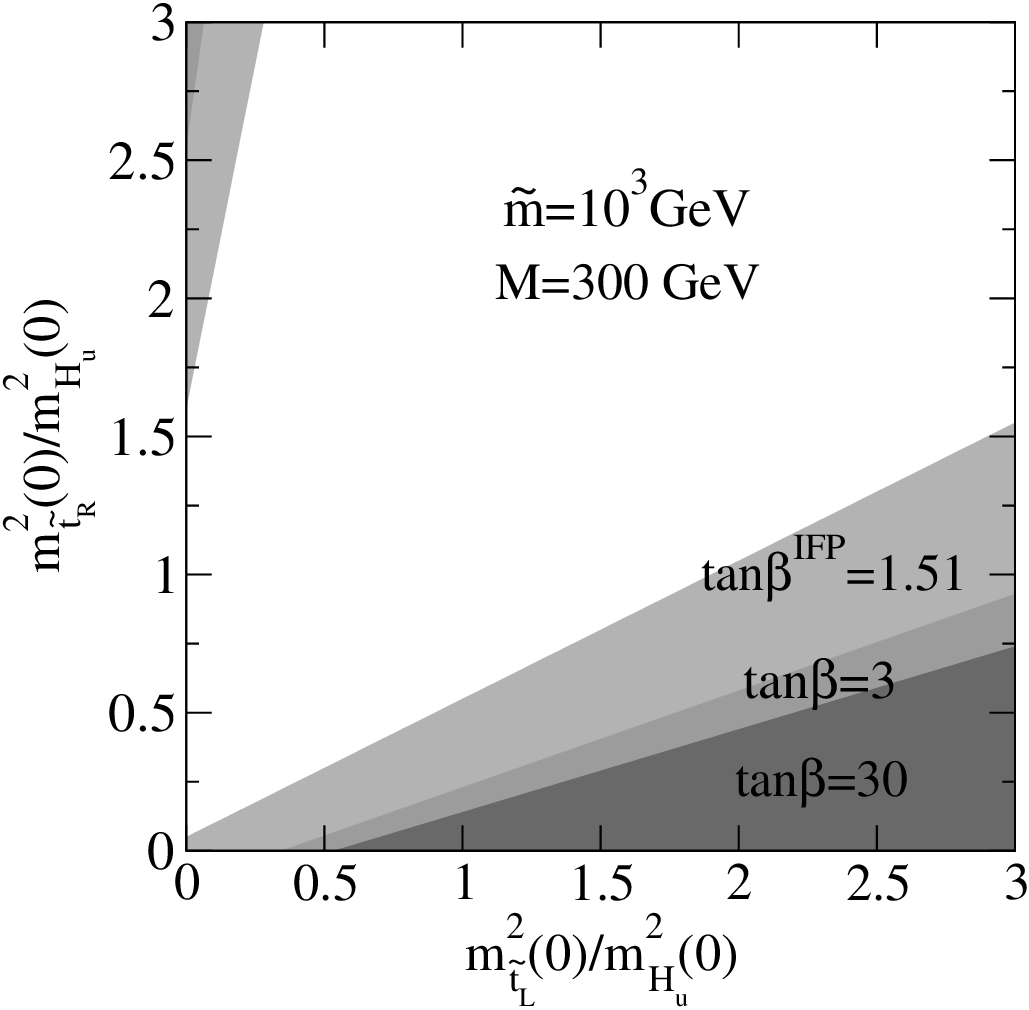}}
  }
  \caption{Allowed regions in the stop parameter space
from imposing the requirement of charge and colour conservation
for different values of $\tan\beta$ and the supersymmetry breaking
mass scale, $\widetilde m$. The lower right plot corresponds
to a situation approaching the conventional MSSM scenario, where
the gaugino masses, $M$, start to be relevant for the analysis.}
  \label{fig-mtilde}
\end{figure}

As $\tan\beta$ increases and one gets away from the
top infrared fixed point, the top Yukawa coupling becomes smaller and 
the evolution of the squark masses squared towards negative
values slows down. Therefore, the constraints on the
squark parameter space that follow from the condition of charge and
colour conservation relax.
This is illustrated in Fig.\ref{fig-mtilde} for different
values of $\tan\beta$ and the soft supersymmetry
breaking scale, $\widetilde m$.
Notice that for a fixed $\tan\beta$ the constraints 
are more restrictive as $\widetilde m$ becomes smaller. 
The reason is that the squark masses are 
running towards negative vales for longer,
so it is easier to develop tachyonic masses.
This behaviour holds as long
as the gaugino masses are much smaller than $\widetilde m$.
For values of $\widetilde m$ close to the electroweak scale, 
{\it i.e.} the standard low energy supersymmetry breaking scenario,
gaugino masses (particularly the gluino mass) can be large 
enough to stop the running
of the squark masses squared towards negative values.
One can estimate the size of the gaugino masses 
and $\widetilde m$ at which this
happens from the full solution to eq.(\ref{RGEs}).
Assuming for simplicity that the trilinear soft terms vanish at $M_P$, 
one obtains:
\bea
m^2_{\widetilde t_L}(t)&=& m^2_{\widetilde t_L}(0) +
M^2\left(\frac{1}{3}e(t)+\frac{8}{3}\frac{\alpha_3(0)}{4\pi} f_3(t)+
\frac{\alpha_2(0)}{4\pi} f_2(t)
-\frac{1}{9}\frac{\alpha_1(0)}{4\pi} f_1(t)\right) \nonumber \\
&& -\frac{h_t^2(0)}{(4 \pi)^2} \frac{F(t)}{D(t)}
( m_{H_u}^2(0)+m_{\widetilde t_R}^2(0)+m_{\widetilde t_L}^2(0)), \nonumber \\
m^2_{\widetilde t_R}(t)&=& m^2_{\widetilde t_R}(0) +
M^2\left(\frac{2}{3}e(t)-H_8(t)\right) -
2\frac{h_t^2(0)}{(4 \pi)^2} \frac{F(t)}{D(t)}
( m_{H_u}^2(0)+m_{\widetilde t_R}^2(0)+m_{\widetilde t_L}^2(0)),
\eea
where $e,f_{1,2,3}$ and $H_8$ are known functions, independent of
$\tan\beta$, that can be found in the Appendix B of ref.\cite{Ibanez:1984vq}.
For $\widetilde m=10^3$GeV, the result can be approximated by
\bea
&&m_{\widetilde t_L}^2(t) \simeq  m_{\widetilde t_L}^2(0) 
-\frac{1.7h_t^2(0) }{1+10.4h_t^2(0)}
( m_{H_u}^2(0)+m_{\widetilde t_R}^2(0)+m_{\widetilde t_L}^2(0))  + 4.7 M^2,
\nonumber \\
&&m_{\widetilde t_R}^2(t) \simeq  m_{\widetilde t_R}^2(0) 
-\frac{3.5h_t^2(0) }{1+10.4h_t^2(0)}
(m_{H_u}^2(0)+m_{\widetilde t_R}^2(0)+m_{\widetilde t_L}^2(0)) +3.5 M^2.
\label{solution-with-M}
\eea
In Fig.\ref{fig-mtilde}, lower right plot we show the allowed parameter
space for the case with universal gaugino masses of 
$M=300$ GeV. One finds that a larger
region is now allowed, even for the infrared fixed point
scenario.
For $\widetilde m=M$, the whole parameter space in the plots, 
$0\leq m_{\widetilde t_L,t_R}^2(0) \leq
3 m_{H_u}^2(0)$, becomes allowed for any value of $\tan\beta$.
 
%%%%%%%%%%%%%%%%%%%%%%%%%%%%%%%%%%%%%%%%%%%%%%%%%%%%%%%%%%%%
%%%%%%%%%%%%%%%%%%%%%%%%%%%%%%%%%%%%%%%%%%%%%%%%%%%%%%%%%%%

Therefore, from the point of view of charge and colour breaking, 
scenarios of split supersymmetry with intermediate values 
for the soft masses ($\sim10^6-10^9$ GeV) are disfavoured
with respect to the conventional MSSM scenario with electroweak 
soft masses ($\sim 10^3$ GeV), where the gluinos 
protect the stop masses from becoming tachyonic. Scenarios
with very large soft masses ($\sim10^{12}$ GeV) are also
favoured, since the squark masses are running over
a smaller energy range, and the renormalization effects
are normally not large enough to generate tachyonic masses.
This value is close to the upper bound on the supersymmetry
breaking scale in split supersymmetry of ${\cal O}(10^{13}$GeV)
for a 1 TeV gluino, coming from negative searches of abnormally heavy isotopes 
\cite{Arkani-Hamed:2004fb}.
On the other hand, when the up-type Higgs mass squared is 
already negative at $M_P$, large regions of the parameter 
space also become allowed.

\begin{figure}[t!]%%%%%%%%%%%%%%%%%%%%%%%%%%%%%%%%%%%%%%%%%%%
  %%%
  \centerline{
    \scalebox{0.45}{\includegraphics{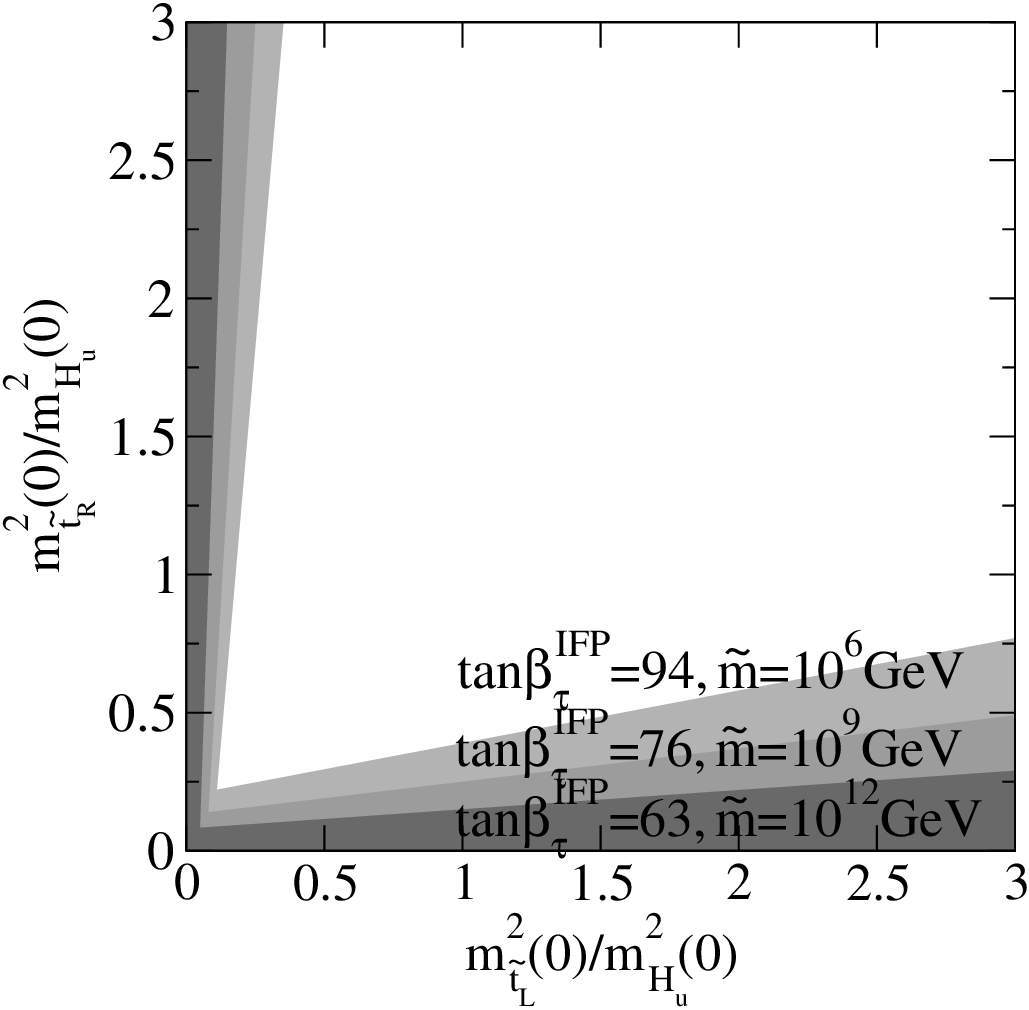}}~~~
    \scalebox{0.45}{\includegraphics{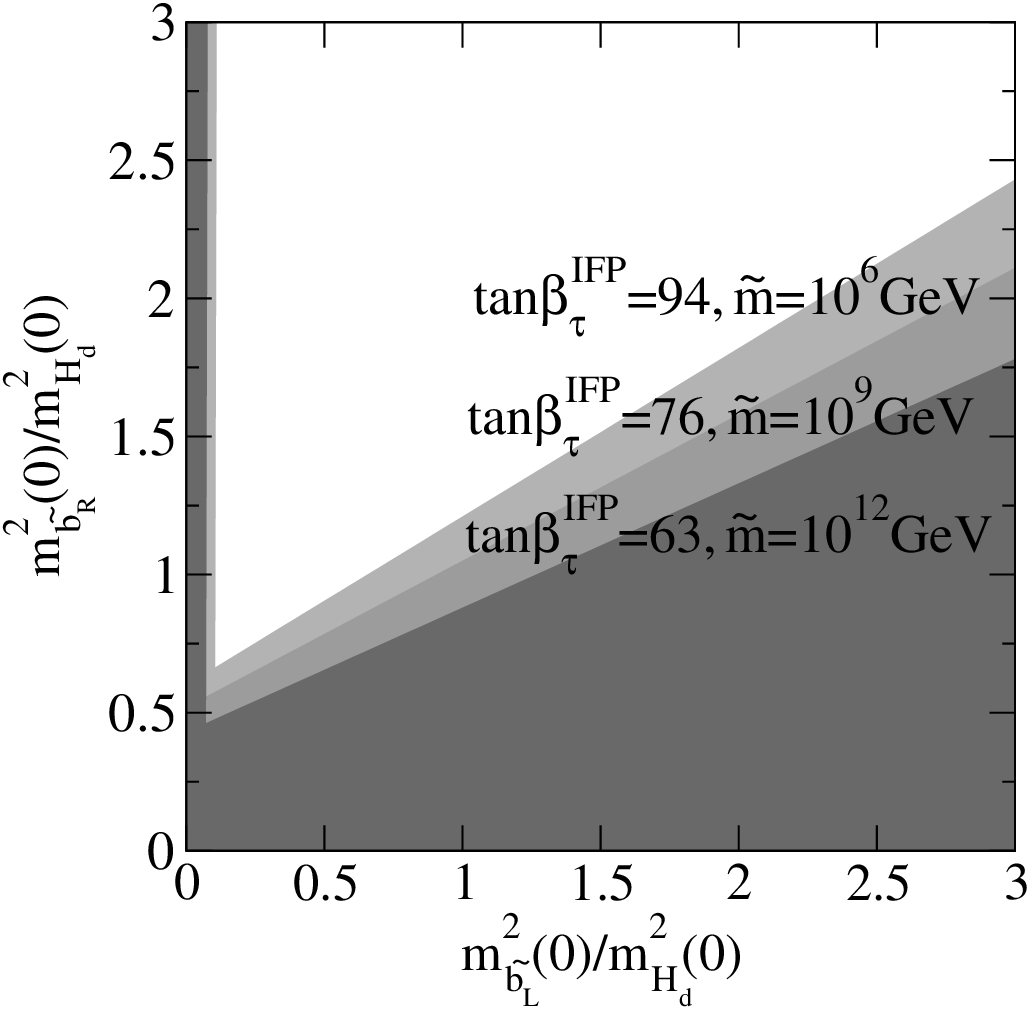}}
  }
  \caption{Allowed region in the stop (left plot) and 
sbottom (right plot) parameter spaces
from imposing the requirement of charge and colour conservation,
in the infrared fixed point limit for the tau Yukawa coupling and
for different values of the SUSY breaking mass scale, $\widetilde m$.}
  \label{fig-stop-sbottom}
\end{figure}

Let us comment now on the situation in which $\tan\beta$ is large. 
Whereas the top Yukawa 
coupling does not change substantially as $\tan\beta$ increases,
the bottom and tau Yukawa coupling indeed do, and their
effects have to be taken into account. 
The allowed range for $\tan\beta$ is limited from above by
the appearance of a Landau pole for the tau Yukawa coupling,
which occurs at $\tan\beta\simeq94$, 76 and 63 for
$\widetilde m = 10^{12}$, $10^9$ and $10^6$ GeV, respectively.
In this range, the bottom Yukawa coupling always remains perturbative
until $M_P$, being the corresponding values  at 
the cut-off scale $h_b(0)\simeq 2.8$,
3.4 and 6.9 respectively (notice that in split supersymmetry
the prediction for bottom-tau unification is lost).
In this regime, left-handed stop masses squared are driven to negative
values faster than for intermediate values of $\tan\beta$,
while the running of the right-handed stops is not modified substantially.
The effect of the  bottom Yukawa coupling
on the left-handed stop mass is not
very important numerically, and the 
allowed region in the stop parameter space is 
similar to the case with intermediate values of 
$\tan\beta$, as can be realized
from Fig.\ref{fig-stop-sbottom}, left plot, where we show the
allowed region in the stop parameter space for different values
of the soft SUSY breaking scale, $\widetilde m$, and for the value
of $\tan\beta$ that corresponds to the infrared fixed point
limit for the tau Yukawa coupling, $\tan\beta_{\tau}^{{\rm IFP}}$.
%%%%%%%%%%%%%%%%%%%%%%%%%%%%%%%%%%%%%%%%%%%%%%%%%%%%%%%%%%%%
%%%%%%%%%%%%%%%%%%%%%%%%%%%%%%%%%%%%%%%%%%%%%%%%%%%%%%%%%%%

The large bottom and tau Yukawa couplings 
could drive the sbottom and stau masses squared negative,
giving rise to constraints on the sbottom and stau parameter
spaces. Despite the bottom Yukawa coupling never reaches the
Landau pole, it can be large enough to induce radiatively
tachyonic masses for the sbottoms, particularly for the values
of $\tan\beta$ corresponding to the infrared fixed point limit
for the tau Yukawa coupling. This is illustrated in 
Fig.\ref{fig-stop-sbottom}, right plot, where we show
the allowed region in the sbottom parameter space in this limit.
On the other hand, since the tau Yukawa coupling
can become very large, the constraints for the stau parameter space can be 
stronger. In the infrared fixed point limit for the tau Yukawa coupling,
the constraints read
\bea
&&m_{\widetilde e}^2(0)-m^2_{H_d}(0)-m^2_{\widetilde L}(0) >0, \nonumber \\
&&3 m_{\widetilde L}^2(0)-m^2_{H_d}(0)-m^2_{\widetilde e}(0) >0. 
\eea
%%%%%%%%%%%%%%%%%%%%%%%%%%%%%%%%%%%%%%%%%%%%%%%%%%%%%%%%%%%%
\begin{figure}[t!]%%%%%%%%%%%%%%%%%%%%%%%%%%%%%%%%%%%%%%%%%%%
  %%%
  \centerline{
    \scalebox{0.45}{\includegraphics{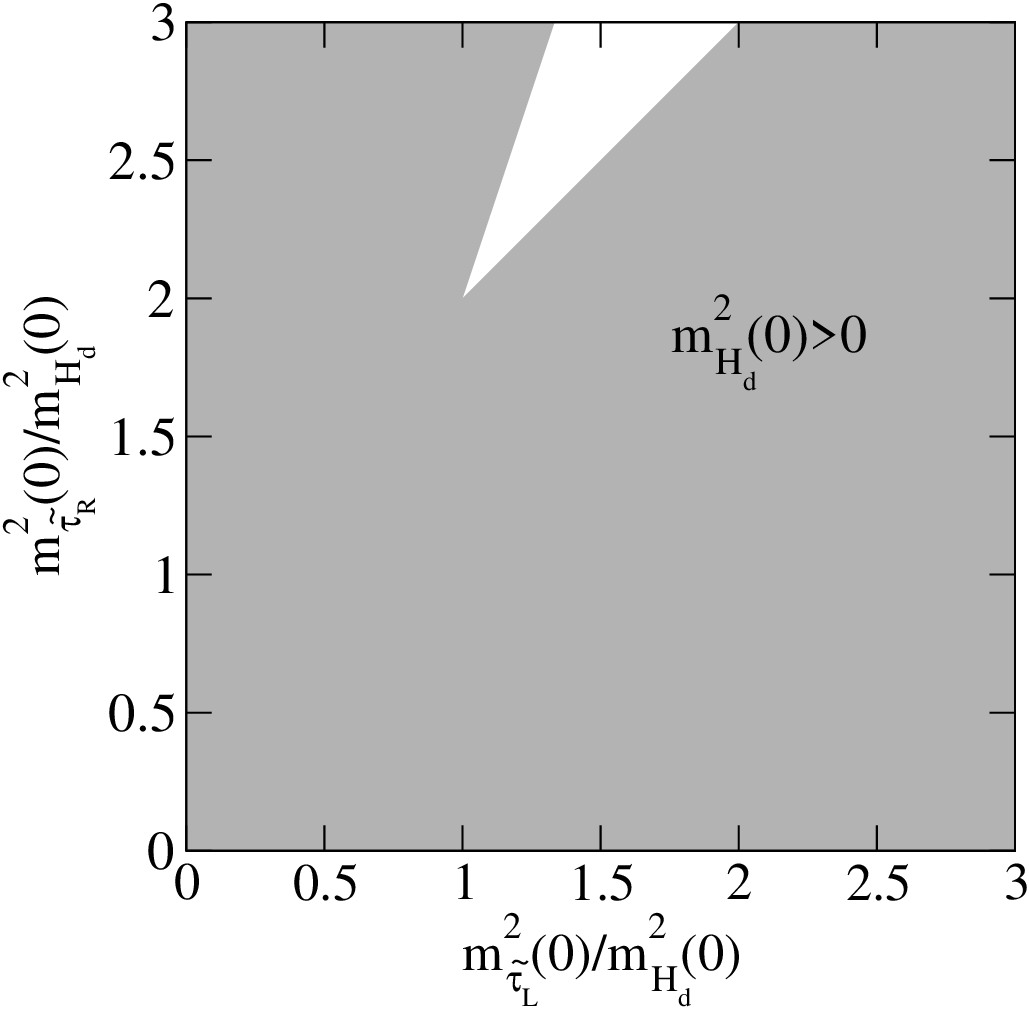}}~~~
    \scalebox{0.45}{\includegraphics{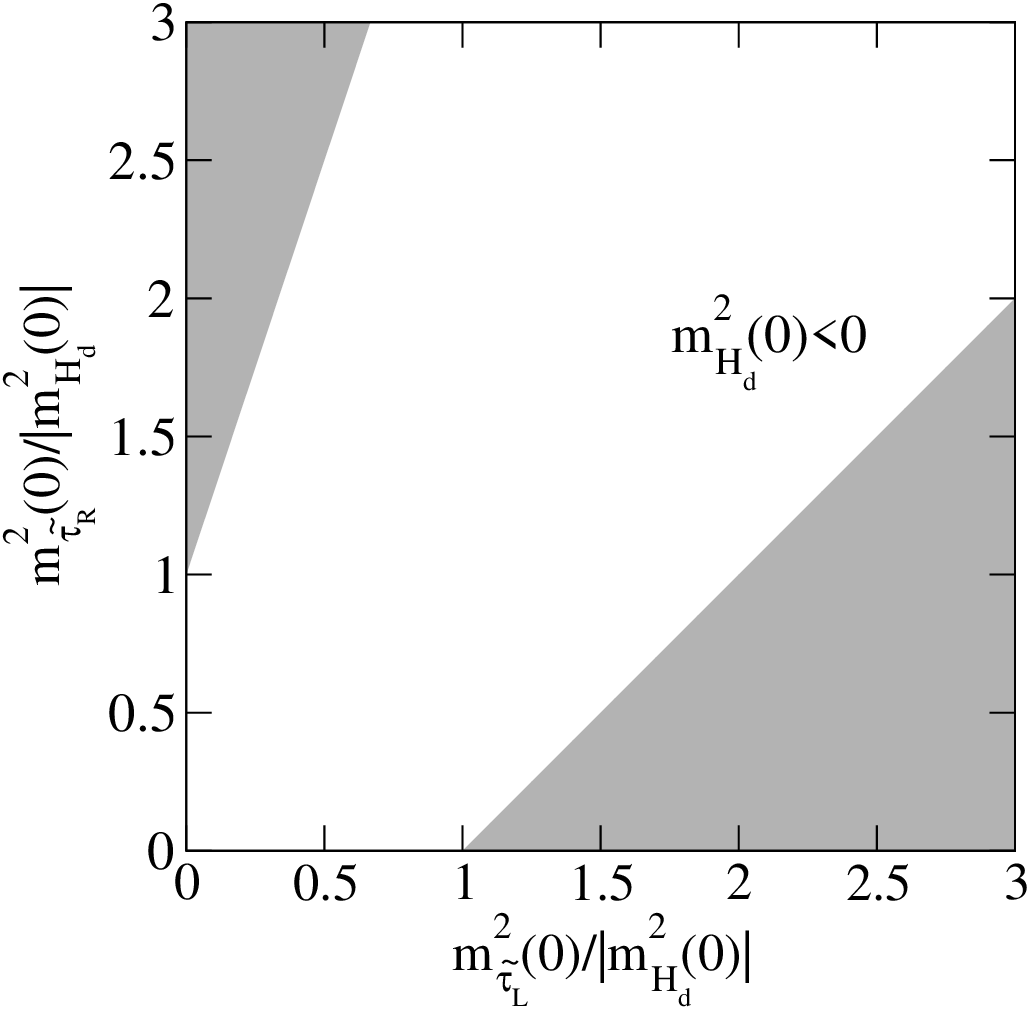}}
  }
  \caption{Allowed region in the stau parameter space
from imposing the requirement of charge conservation,
in the infrared fixed point limit for the tau Yukawa coupling.
In the left plot, it is shown the case in which the down-type
Higgs mass squared is positive at the cut-off scale, whereas in
the right plot we assume that it is negative.}
  \label{fig-IFPtau}
\end{figure}
%%%%%%%%%%%%%%%%%%%%%%%%%%%%%%%%%%%%%%%%%%%%%%%%%%%%%%%%%%%
As can be realized from fig.\ref{fig-IFPtau}, the constraints
on the parameter space from requiring positive stau masses squared
at the decoupling scale are very strong. 
However, these constraints relax as $\tan\beta$ decreases and
practically disappear for small values of $\tan\beta$.
If the particle content of the MSSM is extended with right-handed
neutrinos in order to give masses to neutrinos, 
then the right-handed neutrino Yukawa couplings would
also contribute to the running of the stau at energies larger
than the decoupling scale of the right-handed neutrinos. If the 
neutrino Yukawa couplings are of order one,  
the running of the 
stau mass squared towards negative values would
be considerably accelerated, and
this would translate into stronger bounds on the parameter
space, particularly for small values of $\tan\beta$, for
which the tau Yukawa coupling is small.

\section{Discussion}

We have shown that a split spectrum in supersymmetric
theories is not enough to guarantee the phenomenological
viability of a particular model. Although the decoupling
of squarks and sleptons guarantees the suppression of
flavour changing neutral currents, CP violating 
effects\footnote{Although the contributions 
from the scalar superpartners
to the electric dipole moments are very suppressed, 
the phases in the chargino and neutralino sectors
can propagate at two loops to the Standard Model 
fermions, giving rise to contributions to 
fermion electric dipole moments that could be 
at the reach of future experiments 
\cite{Arkani-Hamed:2004yi,EDM}.}
and proton decay, it does not guarantee that the vacuum
is going to conserve electric charge and colour. 
When the scalar mass spectrum is non-universal, stops could
acquire radiatively tachyonic masses, which is clearly
an undesirable feature. It is important to
stress that despite universality of the scalar
masses is a common assumption in analyses of the supersymmetry
parameter space, this situation is rather exceptional
when constructing supersymmetric models. Most
explicit scenarios of supersymmetry breaking 
predict non-universal scalar masses, where the
analysis presented in this note is particularly
relevant.

Models with split supersymmetry probably require
D-term supersymmetry breaking. Although it is possible
to obtain a split spectrum breaking supersymmetry
giving a vacuum expectation value to an F-term,
this also breaks spontaneously the R-symmetry, 
usually generating gaugino masses and trilinear 
terms of the same order of the scalar masses. 
In contrast, D-term supersymmetry 
breaking does not lead to R-symmetry breaking.
This renders vanishing gaugino masses and
trilinear soft terms at lowest order, which constitutes 
an essential feature of split supersymmetry 
(to generate them it is necessary to add 
non-renormalizable operators in the K\"ahler potential). 
If supersymmetry is indeed broken by the vacuum
expectation value of a D-term, the soft scalar masses
are determined by the charges of the particles
under a particular gauge group, normally generating a
non-universal spectrum. Needless to say, the charges 
have to be arranged in such a way that tachyonic masses 
are not arising already at tree level 
(with the exception of the higgses, possibility that would lead
to the breaking of the electroweak symmetry already
at a high-energy scale).

Some models with split supersymmetry have been constructed
along this lines. The model of Babu, Enkhbat 
and Mukhopadhyaya \cite{Babu:2005ui} utilizes an anomalous 
$U(1)$ symmetry and a gaugino condensate
to trigger supersymmetry breaking\footnote{Scenarios
of split supersymmetry with Fayet-Iliopoulos D-terms were
considered before in \cite{Kors:2004hz}.}. The anomalous $U(1)$ symmetry
is horizontal, thus providing also an explanation for the quark and 
lepton masses and mixing angles. In this model, the different
charges of the particles under the anomalous $U(1)$ generates
a scalar mass spectrum that is non-universal. There are however some
massless fields in the limit of global supersymmetry,
namely the two Higgs doublets and the 
third generation of the 10-plet of $SU(5)$ 
(they have to be neutral under the anomalous $U(1)$ in order 
to reproduce the observed masses and mixing angles).
These particles acquire masses of the order of the
gravitino mass through supergravity corrections,
and could also be non-universal if the K\"ahler 
potential is non-minimal.

A different class of models was proposed by
Antoniadis and Dimopoulos \cite{Antoniadis:2004dt},
and is based on type I string theory with internally
magnetized D9 branes \footnote{This is not the only
possibility and it is also possible to obtain a split
spectrum by intersecting D6 branes\cite{Kokorelis:2004dc}.}.
In the T-dual picture, the model
can be described by intersecting branes, with broken 
supersymmetry when the branes are intersecting at arbitrary angles.
In the four dimensional
effective theory, the breaking of supersymmetry can be
interpreted as a D-term supersymmetry breaking. Therefore,
the different scalar fields will acquire masses 
that depend on the D-terms associated to the 
different $U(1)$s of the theory (or in the dual picture,
on the magnetic fields in the compact dimensions).
In general, the D-terms are different for 
each $U(1)$, leading to a non-universal spectrum.

Finally, we would like to mention that 
in scenarios with split supersymmetry there exists
a second threat for charge and colour conservation,
namely the appearance of unbounded from below
directions in the effective potential,
that could lead to deep charge and colour breaking
minima after including radiative corrections
\cite{Casas:1995pd}. Again, whether 
these directions appear or not depends on the spectrum on squarks
and sleptons, but not on the scale of supersymmetry breaking,
therefore they could also arise in models with split 
supersymmetry\footnote{In the conventional MSSM with low
energy supersymmetry breaking,
there are also charge and colour breaking minima
appearing at tree level, due to the negative contribution
to the effective potential from the trilinear soft terms. The
smallness of the trilinear terms in split supersymmetry guarantees
that these minima are not present.}. Nevertheless,
it has been argued in \cite{Abel:1998ie} that even if
they appear, the decay rate
of the metastable electroweak minimum into the global minimum
is very suppressed, so that the lifetime of the metastable
minimum is usually longer than the age of the Universe.
In consequence, the constraints that would follow
from requiring the absence of unbounded from below
directions could be avoided if one accepts that
our electroweak vacuum is a metastable minimum with a small
cosmological constant. This hypothesis might look dubious,
however, in the spirit of split supersymmetry,
it could be justified by some anthropic principle.
Hence, the only constraints on the scalar spectrum that are robust
are the ones discussed in this note, namely the possibility of
radiative generation of tachyonic squark masses.

\section{Conclusions}

Split supersymmetry is a daring proposal to solve
all the problems of the Supersymmetric Standard Model,
while preserving the successes. The decoupling of
the scalar particles, except for the 
Standard Model Higgs, suppresses flavour changing 
neutral currents, electric dipole moments
and the rates for proton decay. Besides, imposing 
global symmetries to keep the gauginos and higgsinos
light, preserves the nice features of gauge unification
and the neutralino as a dark matter candidate. 

If this scenario is realized in nature, squarks and sleptons
would not have any observable effect in low
energy processes, neither at tree level nor at the radiative
level. However, we have remarked in this
note that their spectrum is not totally unconstrained,
even though squarks and sleptons are completely
decoupled at low energies, and we have discussed the
implications for building models with split supersymmetry. 
We have shown that the
structure of the vacuum depends crucially on
the spectrum of squarks and sleptons, and
we have derived the constraints that follow
from the requirement of charge and colour conservation.
To be precise, certain patterns of supersymmetry breaking could
induce radiatively tachyonic masses for the stops
(and for the staus when $\tan\beta$ is large),
thus breaking electric charge and colour. In particular,
models with an intermediate supersymmetry breaking
scale ($\widetilde m \sim 10^6-10^9$ GeV) are disfavoured with respect
to the conventional MSSM scenario with low energy supersymmetry
breaking ($\widetilde m \sim 10^3$ GeV). We have also stressed
that models with split supersymmetry probably
require D-term supersymmetry breaking, leading in general
to a non-universal spectrum where the constraints
presented in this note are potentially dangerous.

\section*{Acknowledgments}
I would like to thank Alberto Casas, Anamar\'{\i}a Font, Gian Giudice,
Andrea Romanino, Angel Uranga and Sudhir Vempati for very 
interesting discussions, and to the CERN Theory Division 
for hospitality during the last
stages of this work.

%%%%%%%%%%%%%%%%%%%%%%%%%%%%%%%%%%%%%%%%%%%%%%%%%%%%%%%%%%%%%%%%%%%

\end{document}